\documentclass[aps,prl,twocolumn,superscriptaddress,longbibliography,nobalancelastpage]{revtex4-2}

\usepackage{graphicx}
\usepackage{amsmath}
\usepackage{amssymb}
\usepackage{color}
\usepackage[normalem]{ulem}
\usepackage[utf8]{inputenc}
\usepackage[colorlinks = true,
            allcolors = {Blue}]{hyperref}

\usepackage[dvipsnames]{xcolor}
\usepackage{comment}
\newcommand{\review}[1]{#1}

\newcommand{\op}[1]{\hat{#1}}
\newcommand{\abs}[1]{\ensuremath{|#1|}}
\newcommand{\ket}[1]{| {#1} \rangle }
\newcommand{\bra}[1]{\langle {#1} | }
\newcommand{\expect}[1]{\langle #1 \rangle}
\newcommand{\floor}[1]{\lfloor #1 \rfloor}
\newcommand{\fractional}[1]{\mathrm{frac}(#1)}

\begin{document}

\title{Efficient exploration of Hamiltonian parameter space for optimal control of non-Markovian open quantum systems}
\author{Gerald E. Fux}
\affiliation{SUPA, School of Physics and Astronomy, University of St Andrews, St Andrews, KY16 9SS, United Kingdom}
\author{Eoin P. Butler}
\affiliation{School of Physics, Trinity College Dublin, College Green, Dublin 2, Ireland}
\author{Paul R. Eastham}
\affiliation{School of Physics, Trinity College Dublin, College Green, Dublin 2, Ireland}
\author{Brendon W. Lovett}
\affiliation{SUPA, School of Physics and Astronomy, University of St Andrews, St Andrews, KY16 9SS, United Kingdom}
\author{Jonathan Keeling}
\affiliation{SUPA, School of Physics and Astronomy, University of St Andrews, St Andrews, KY16 9SS, United Kingdom}

\date{\today}

\begin{abstract}
	We present a general method to efficiently design \review{optimal control sequences for non-Markovian open quantum systems}, and illustrate it by optimizing the shape of a laser pulse to prepare a quantum dot in a specific state. \review{The optimization of control procedures for quantum systems} with strong coupling to structured environments---where time-local descriptions fail---is a computationally challenging task. We modify the numerically exact time evolving matrix product operator (TEMPO) method, such that it allows the repeated computation of the time evolution of the reduced system density matrix for various sets of control parameters at very low computational cost. This method is potentially useful for studying numerous \review{optimal control} problems, in particular in solid state quantum devices where the coupling to vibrational modes is typically strong.
\end{abstract}

\maketitle


One of the main challenges in current quantum technology is to avoid or mitigate decoherence. 
\review{The field of ``quantum optimal control theory"~\cite{Shapiro2012,Torrontegui2013,Glaser2015,Koch2016,Koch2019} seeks to address this challenge by searching for classical control sequences on quantum systems to} achieve the highest fidelity of a process for a given experimental setup. 
For this to be successful, it is necessary to be able to accurately compute the dynamics of the system under the influence of the environment. The majority of research on optimal control for open quantum systems has been carried out in the Markovian limit, where one assumes a weak system-environment coupling and environment correlations that are short compared to the timescale of the system evolution~\cite{Breuer2002,Chirolli2008}. However, in many solid-state devices and other systems this assumption breaks down~\cite{WilsonRae2002,Galland2008,Ramsay2010,Roy2011,Luker2012} so one cannot use simple time-local density matrix equations of motion. In addition to non-Markovianity being common, it can be desirable~\cite{Rebentrost2009,Schmidt2011,  Hwang2012, Floether2012,Reich2015,Addis2016,Puthumpally-Joseph2018,Mangaud2018, Goerz2018,Fischer2019,Mirkin2019,Alipour2020}: it has been shown that non-Markovianity of open quantum systems can lead to higher fidelity of quantum operations due to the possible information backflow from the environment to the system. The simulation of general non-Markovian open quantum systems is, however, a computationally challenging task, which hampers progress on the design of optimal control procedures.

There exist a number of available methods applicable to simulating specific non-Markovian scenarios~\cite{Tanimura1989, Makri1995, Makri1995a,  Prior2010, Chin2010,Cerrillo2014, Iles-Smith2014,  Tamascelli2017,  Tamascelli2018,  Somoza2019, Mascherpa2019, Brenes2019, Tanimura2020} (see~\cite{DeVega2017} for a review of some of these). One approach is to extend the notion of the system, in cases where the environment can be modeled as extra ``noise qubits'' which couple strongly to the system and weakly to some Markovian environment~\cite{Rebentrost2009, Reich2015, Fischer2019}. This is done systematically in the time evolving density operator with orthogonal polynomials (TEDOPA)~\cite{Prior2010, Chin2010} method, which maps the environment to a chain of coupled sites.
Instead of augmenting the system space, one can write coupled hierarchical equations of motion (HEOM)~\cite{Tanimura1989, Puthumpally-Joseph2018, Mangaud2018, Tanimura2020}; this performs well for spectral densities that are well-approximated by a small number of Lorentzians.  Most relevant to this Letter are methods based on an augmented density tensor, capturing the system history, specifically the quasi adiabatic path integral (QUAPI)~\cite{Makri1995, Makri1995a}. When considering optimal control applications,  a major impediment to all of these methods is that they are computationally intensive and the entire calculation needs to be repeated for each set of control parameters.  This makes numerical optimization unfeasibly costly.

\begin{figure}[b]
	\includegraphics[width=0.45\textwidth]{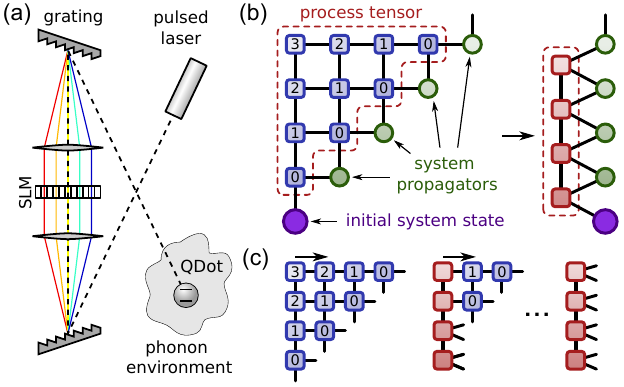}
	\caption{\label{fig:pulse-shaper-process-tensor} (a) Sketch of the experimental setup to drive a quantum dot (QDot) with a shaped laser pulse. The pulse form can be modified with a spatial light modulator (SLM). (b) The TEMPO tensor network for four time steps. (c) Contraction scheme to obtain the process tensor in MPS form. \review{The numbers $n\in\mathbb{N}$ on the blue boxes in (b) and (c) label the influence functionals $I_n$ that account for the influence of the system on itself via the environment at a time delay of $n\,\delta t$.}}
\end{figure}

In this Letter we present a general method to efficiently design \review{classical control procedures for non-Markovian open quantum systems}. A crucial step is a recasting of the TEMPO method~\cite{Strathearn2018, Strathearn2019} within the framework of process tensors 
introduced by Pollock \textit{et al.} in Ref.~\cite{Pollock2018}.  With the resulting process tensor TEMPO (PT-TEMPO) method~\cite{TimeEvolvingMPO} one can perform the bulk of the computation independent of the system control parameters. This enables us to repeatedly find the system density matrix time evolution for various sets of control parameters at very low computational cost. We can use this to optimize \review{classical control procedures} with respect to any chosen aspect of the system evolution, taking full account of non-Markovian effects. To demonstrate this method we apply it to a quantum dot that is driven by a shaped laser pulse and strongly coupled to a super-Ohmic phonon environment (see Fig.~\ref{fig:pulse-shaper-process-tensor}a for a sketch of the experimental setup we model). We show that our method can explore a thirty-five dimensional space of control parameters, and find optimized pulses for state preparation in an ensemble of five qubits of differing energies.


\textit{TEMPO and process tensors}---
The most general scenario that we consider is a small system coupled to a bosonic environment with a total Hamiltonian of the form
\begin{equation}
	\label{eq:model-hamiltonian}
	\op{H} = \op{H}_\mathrm{S}(t,\{\review{c_n}\}) + \sum_{k=0}^{\infty} \left[ \op{O}_S\! \left( g_k \op{b}_k^\dagger\! +\! g_k^* \op{b}_k \right) \! + \! \omega_k \op{b}_k^\dagger \op{b}_k \right] \mathrm{,}
\end{equation}
\review{where the system Hamiltonian $\op{H}_S$ depends on the classical control sequence---which is parametrized by a set of real variables $\{c_n\}$---and is therefore time dependent. The operator $\op{O}_S$ is the system coupling operator, $g_k$ are the coupling constants, $\omega_k$ are the environment mode frequencies, and $\op{b}^{(\dagger)}_k$ are the bosonic environment lowering (raising) operators. The coupling constants and environment mode frequencies are encoded in the spectral density \mbox{$J(\omega) = \sum_{k=0}^{\infty} \abs{g_k}^2 \delta(\omega - \omega_k)$}.}
We assume that at some initial time $t_0$ the total state can be written in a product state $\rho(t_0)=\rho_S(t_0)\otimes\rho_E(t_0)$, where $\rho_E(t_0)$ is a Gaussian state of the bosonic environment (for example a thermal state). We make no assumption on the coupling strength or the total state at any later time.

The TEMPO method is numerically exact and based on Feynman-Vernon influence functionals~\cite{Feynman1963}. Like QUAPI, the TEMPO method utilizes an augmented density tensor (ADT) to encode the system's history and its auto-correlations over time. It employs tensor network methods to compress this ADT in the form of a matrix product state (MPS)~\cite{Orus2014,Cirac2020}, which allows it to include the history over hundreds of time steps. Figure~\ref{fig:pulse-shaper-process-tensor}b exemplifies the TEMPO tensor network for four time steps. Each node of this network represents a tensor and each edge (called \emph{leg}) corresponds to an index. When a leg connects two tensors it signals a summation between them.

\review{The TEMPO method uses a Suzuki-Trotter expansion of the total propagator $e^{-i \op{H} \delta t}$ into propagators $e^{-i \op{H}_\mathrm{S} \delta t} e^{-i \op{H}_\mathrm{E} \delta t}$ of the pure system part $H_\mathrm{S}(t)$ and the remainder $H_\mathrm{E}$, plus higher order terms $\mathcal{O}(\delta t^2)$. The time step $\delta t$ must be chosen small enough that these higher-order terms can be neglected.}
The tensor network underlying \review{this method} works in Liouville space, which means that density matrices are represented as vectors and the maps between them as matrices (so called \emph{super-operators}). The \review{purple} circle in Fig.~\ref{fig:pulse-shaper-process-tensor}b is the vectorized initial system state $\rho_S(t_0)$ and the green circles are the system propagators for a single time step \mbox{$\mathcal{U}_m= \exp\left( \mathcal{L}_S(t_m+\delta t/2) \delta t \right)$} at times \mbox{$t_m = t_0 + m \delta t$} with the system Liouvillian super-operator \mbox{$\mathcal{L}_S(t) = -i [\op{H}_S(t), \cdot] $}. The blue squares in Figs.~\ref{fig:pulse-shaper-process-tensor}b~and~\ref{fig:pulse-shaper-process-tensor}c are the Feynman-Vernon influence functionals $I_n$, which can be directly constructed \review{from} the coupling operator $\op{O}_S$, the spectral density $J(\omega)$ and the initial environment state $\rho_E(t_0)$. An influence functional $I_n$ quantifies how the system evolution at some time $t_m$ is influenced by the state of the system at the earlier time $t_{m-n}$, thus allowing for a non-Markovian dynamics of the system. \review{The influence functionals depend only on the time difference $n\,\delta t$ between $t_m$ and $t_{m-n}$, because the environment is time translational invariant.} Because the individual Feynman-Vernon influence functionals in the network are derived from the corresponding time intervals in the environment auto-correlation function, the TEMPO method performs best when this function is smooth and decays to zero within some finite time. For more details on this method \review{and the precise form of the $I_n$ tensors,} see~Refs.~\cite{Strathearn2018,Strathearn2019}.

The crucial point we make use of is that since most of the TEMPO tensor network consists solely of influence functionals, which do not depend on the system Hamiltonian or the initial system state (see red dashed region in Fig.~\ref{fig:pulse-shaper-process-tensor}b), we may perform the bulk of the computation---contraction of the tensor network---before specifying the system Hamiltonian.  This provides an efficient method enabling optimization over \review{system} Hamiltonian parameters.

As realized by J{\o}rgensen and Pollock\review{,} the TEMPO network can be contracted to yield a \emph{process tensor}~\cite{Pollock2018,Jorgensen2019}. The process tensor framework takes an operational approach to characterize non-Markovian open quantum systems\review{, by considering a finite set of interventions. Its central object---the process tensor---has two legs for each intervention and encodes the outcome of every possible sequence.} It has been used to resolve common misconceptions on implications of completely positive divisibility~\cite{Milz2019} and it gives a natural definition of quantum non-Markovianity that coincides with the classical definition in the classical limit~\cite{Pollock2018a}. Here we show that in addition to these conceptional advantages it also has computational benefits. 
\review{If one chooses a small enough time step such that a Suzuki-Trotter expansion as described above is a good approximation, then the system Hamiltonian need not be part of the computation of the process tensor. The red dashed area appearing in the TEMPO tensor network in Fig.~\ref{fig:pulse-shaper-process-tensor}b can be identified as such a process tensor with respect to a total Hamiltonian that excludes the pure system part $\op{H}^\prime = \op{H} - \op{H}_S(t,\{p_n\})$.  Thus, in the language of the process tensor formalism, the red dashed area in Fig.~\ref{fig:pulse-shaper-process-tensor}b is a process tensor and the pure system propagators are a set of interventions.}

To optimize the control of a non-Markovian open quantum system, we propose to make use of the ideas above and split the computation into two steps. First, we contract the influence functionals column by column as \review{explained in detail} in~\cite{Jorgensen2019} and depicted schematically in Fig.~\ref{fig:pulse-shaper-process-tensor}c. The result of this computation yields the process tensor in MPS form, which we save. Then, we perform a search by repeatedly inserting different system propagators associated with various sets of control parameters. This allows us to compute the reduced system dynamics at very low computational cost and thus enables us to find \review{control sequence} parameters that optimize the fidelity of the process.


\textit{Application to a quantum dot}---
We demonstrate the performance of the PT-TEMPO approach by applying it to a quantum dot that is strongly coupled to its phonon environment and driven by a configurable laser pulse.  We aim to drive the quantum dot into a coherent superposition within a few picoseconds. Modelling this is challenging because the evolution timescale is comparable to the memory time, so non-Markovian effects play an important role.

Figure~\ref{fig:pulse-shaper-process-tensor}a shows a sketch of the experimental setup we model for this purpose. The laser pulse shape is modified with a pair of diffraction gratings, lenses, and a spatial light modulator (SLM). We consider the ground state and the exciton state of the quantum dot and denote them with $\ket{\downarrow}$  and $\ket{\uparrow}$ respectively.  Under the rotating wave approximation the system Hamiltonian (with $\hbar=1$) is
\begin{equation}
	\op{H}_S(t)= \frac{\omega_{\uparrow\downarrow}}{2} \op{\sigma}_z
	+ \frac{\Omega(t)}{2} e^{-i \omega_0 t} \op{\sigma}^+
	+ \frac{\Omega^*(t)}{2} e^{i \omega_0 t} \op{\sigma}^- \mathrm{,}
\end{equation}
where $\Omega(t)$ is the positive frequency part of the classical electrical field amplitude, $\omega_0$ is the laser carrier frequency and $\omega_{\uparrow\downarrow}$ is the exciton energy. Also, $\sigma_z$ is the Pauli matrix, $\sigma^+ = \ket{\uparrow}\bra{\downarrow}$, and $\sigma^- = \ket{\downarrow}\bra{\uparrow}$. 
In addition, the quantum dot couples strongly to its phonon environment with the coupling operator \mbox{$\op{O}_S = \sigma_z / 2$} and a super-Ohmic spectral density \mbox{$J(\omega) = 2 \alpha \omega^3 \omega_c^{-2} \: \exp(-\omega^2/\omega_c^2 )$}, with the unit-less coupling constant $\alpha= 0.126$ and the cut-off frequency \mbox{$\omega_c= 3.04\,\mathrm{ps}^{-1}$}~\cite{Ramsay2010,Luker2012,Eastham2013}. The initial state is assumed to be the product of the quantum dot ground state $\ket{\downarrow}$ and the thermal state of the environment at $T=1\,\mathrm{K}$. We note that the environment auto-correlation function dies off only after about $2.5\,\mathrm{ps}$, which renders a Markovian approach invalid at comparable and shorter timescales.

For convenience, we transform the system Hamiltonian into a frame rotating at the frequency of the optical transition, such that
\begin{equation}
	\label{eq:q-dot-hamiltonian}
	\op{H}_\mathrm{S}(t) = \frac{\mathcal{E}(t)}{2}  \op{\sigma}^+ + \frac{\mathcal{E}^*(t)}{2} \op{\sigma}^- \mathrm{,}
\end{equation}
where $\mathcal{E}(t) = \Omega(t) \exp(-i \Delta t)$ is the positive frequency part of the electric field in the rotating frame, and \mbox{$\Delta = \omega_0 - \omega_{\uparrow\downarrow}$} is the detuning of the carrier frequency of the input pulse with respect to resonance. The input pulse (before it enters the pulse shaper) is assumed to be Gaussian, i.e. \mbox{$\mathcal{E}_\mathrm{in}(t) \propto \tau^{-1} \exp\left(-t^2 / \tau^2\right) \exp\left(-i \Delta t\right)$}, with the input pulse duration $\tau$ assumed to be between 30\,fs and 300\,fs. The pair of diffraction gratings and appropriate lenses enable the spatial separation of the frequency components of the input pulse, with an approximately linear relationship between frequency and position at the SLM. Therefore each SLM pixel modifies a particular frequency range of the pulse. We further assume that the pulse also has a finite spatial width with a Gaussian profile, which results in a finite spot size for each frequency part at the SLM. Given that each SLM pixel can induce a phase shift $\phi_n$ to its corresponding frequency $\Omega_n$ the output pulse $ \mathcal{E}(t) \propto \left(h \ast \mathcal{E}_\mathrm{in}\right)(t) $ is a convolution with the SLM's impulse response function
\begin{equation}
	\label{eq:pulse-shaper-response-function}
	h(t) \propto \mathrm{sinc}\left(\frac{\delta\Omega_p\,t}{2}\right)  e^{-\frac{\delta\Omega_s^2 t^2}{4}} \sum_{n \in \mathrm{pixels}} e^{i (\Omega_n t + \phi_n)} \mathrm{,}
\end{equation}
where $\delta\Omega_p$ is the pixel width and $\delta\Omega_s$ is the spot size in terms of their corresponding frequency range~\cite{Weiner1992, Weiner2000}. We assume 512 SLM pixels centered at the pulse carrier frequency and evenly spaced over a frequency range of $2\pi\times128.0\,\mathrm{ps}^{-1}$. Also, we assume that the spot size of the pulse covers about two pixels, i.e. $\delta\Omega_s=2.0 \times \delta\Omega_p$.

The setup described here leads to 515 experimental parameters to modify the laser pulse, namely, the initial pulse length $\tau$, the initial pulse detuning $\Delta$, the pulse area $\Theta$, and one phase shift $\phi_n$ for each of the 512 SLM pixels. Instead of directly using the 512 parameters on the SLM, we use a continuous phase mask function $f(x)$ on the interval $x \in [-1, 1]$, where $-1$ is mapped to the most red detuned pixel and $1$ is mapped to the most blue detuned pixel of the SLM. Then, the phase shift $\phi_n$ of pixel $n$ is \mbox{$\phi_n = 2\pi \, \fractional{f(x(n)) / 2\pi} \in [0,2\pi) $}, where $x(n) = (n-256)/256$ and $\fractional{y} = y - \floor{y}$ denotes the fractional part.

To study the non-Markovian dynamics of the quantum dot as a function of these experimental parameters we employ the PT-TEMPO method for which we first compute the process tensor. Similar to the conventional TEMPO method, the accuracy of the result as well as the necessary computation time depends on the choice of the simulation parameters. We choose time steps of $10\,\mathrm{fs}$, a memory time of $2.5\,\mathrm{ps}$ and we truncate singular values that are smaller than $10^{-6.5}$ relative to the largest value during the contraction. With this, the computation of the process tensor takes approximately $167\,\mathrm{s}$, which only needs to be calculated once. The application of a system Hamiltonian to this process tensor takes only $1.7\,\mathrm{s}$ on a single core of an Intel I7 (8th Gen) processor. For comparison, a conventional TEMPO computation~\cite{Strathearn2018} leading to a comparable accuracy of the result takes approximately $230\,\mathrm{s}$ for each set of control parameters.

\begin{figure}[b]
	\includegraphics[width=0.45\textwidth]{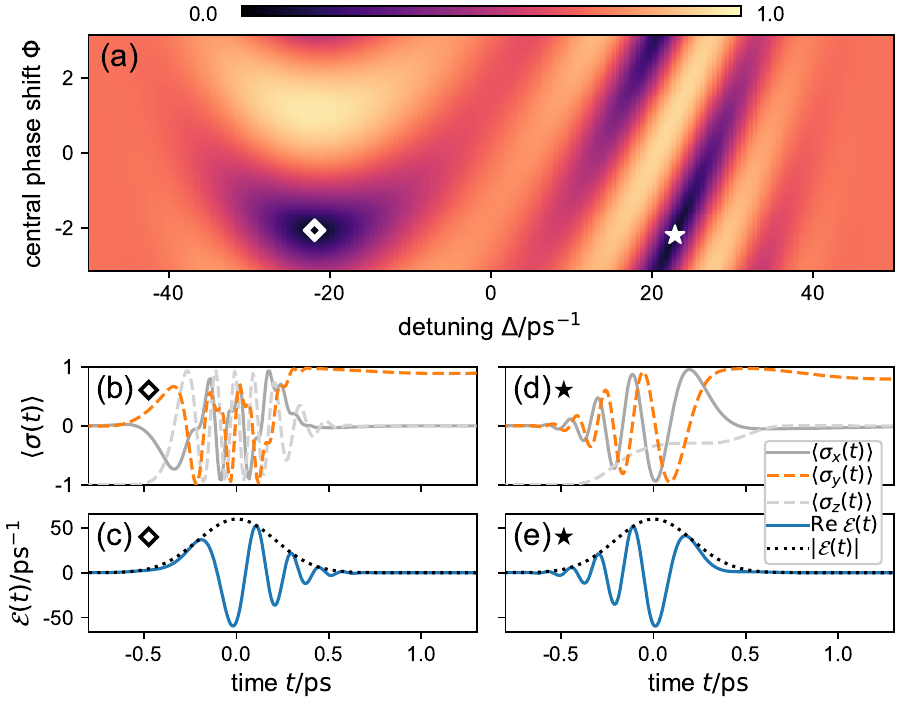}
	\caption{\label{fig:landscape} The dynamics of a quantum dot as a function of the detuning and overall phase of a chirped laser pulse. (a)~A heat map indicating the trace distance of the final state to the target state $\ket{y+}$. (b-e)~Dynamics of the quantum dot and the electric field for the laser pulse parameters marked with the symbols $\diamond$ and $\star$ in (a) respectively.}
\end{figure}
As a first example we apply a laser pulse to drive the quantum dot from its ground state $\ket{\downarrow}$ to the \mbox{$\ket{y+} = (\ket{\uparrow} + i \ket{\downarrow}) /\sqrt{2}$} state. For simplicity we pick a two dimensional parameter space, for which we fix the initial pulse length to $\tau = 50\,\mathrm{fs}$ and the pulse area to $\Theta = 10\,\pi$. We also fix the shape of the phase mask function to a downward facing parabola $f(x) = \Phi - 1300\,x^2$ with a central phase shift $\Phi$. This parabola results in a broadened and chirped output laser pulse that starts blue detuned and ends red detuned with respect to its carrier frequency. The central phase shift induces an overall phase which rotates the x-y coordinate system.
Applying our method we can easily map out the trace distance of the final state to the $\ket{y+}$ target state, as a function of the two open parameters $\Delta \in [-50 , 50]\,\mathrm{ps}^{-1}$ and $\Phi \in [-\pi,\pi]$. Figure~\ref{fig:landscape}a shows the results of $201 \times 81$ full non-Markovian simulations corresponding to the different parameter sets. Employing the PT-TEMPO method the entire computation takes less than 8 hours on a single core of an Intel i7 processor, while it would take approximately 1040 hours or 43 days employing the conventional TEMPO method. We find two local minima on this landscape which are marked by a star and a diamond in Fig.~\ref{fig:landscape}a. The laser pulse that corresponds to the star parameter set is a chirped pulse that starts strongly detuned and finishes on resonance. This can be thought of as an interrupted adiabatic rapid passage, which has the advantage to be almost independent of an inaccurate pulse area, but has the disadvantage to be sensitive towards the detuning of the pulse. The laser pulse that corresponds to the diamond parameters, on the other hand, starts on resonance and ends strongly red detuned. In this case, like a dynamical gate, the fidelity of the protocol is sensitively dependent on the pulse area, but tolerant towards detuning inaccuracies, similar to a simple $\pi/2$-pulse.

\begin{figure}[b]
	\includegraphics[width=0.45\textwidth]{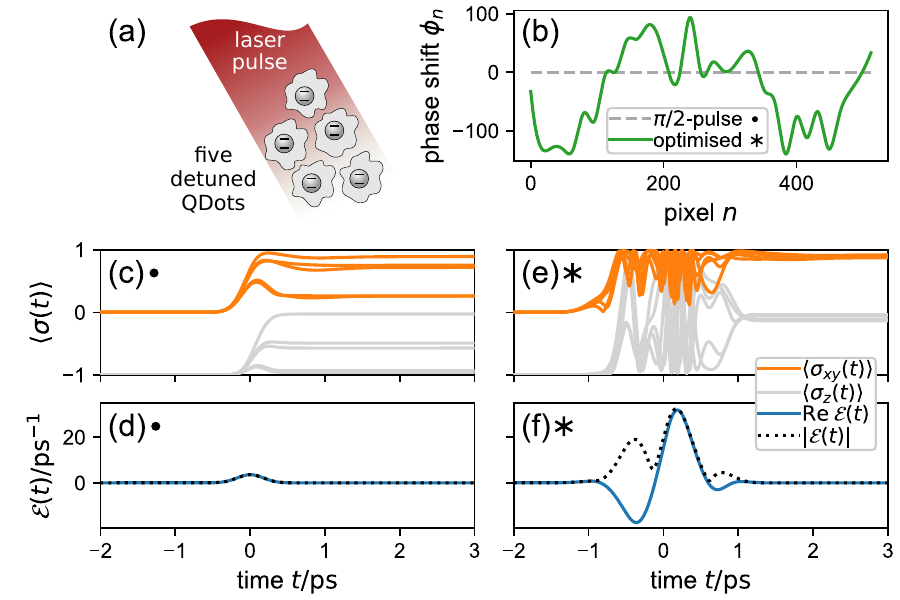}
	\caption{\label{fig:ensemble} Optimization of a laser pulse driving an ensemble of five quantum dots (QDots). (a)~A sketch of the ensemble taking the place of the single QDot in the setup from Fig.~\ref{fig:pulse-shaper-process-tensor}a. (b)~The phase mask function for the $\pi/2$-pulse and the optimal laser pulse. (c-f)~Dynamics of the QDots and the electric field for the $\pi/2$-pulse and the optimal laser pulse denoted with $\bullet$ and $\ast$ respectively. The pulse length of both pulses prior to the pulse shaper is $245\,\mathrm{fs}$, the pulse areas for the initial pulse and the optimized pulse are $0.5\times\pi$ and $7.56\times\pi$ respectively. The plots in (c) and (e) show the expectation values $\expect{\sigma_{xy}(t)} = \sqrt{\expect{\sigma_x(t)}^2 + \expect{\sigma_y(t)}^2}$ for all five QDots.}
\end{figure}
As a second example to demonstrate the performance of the PT-TEMPO method we consider a set of five individually detuned quantum dots (each with their own environment) and aim to find a global optimal laser pulse to simultaneously drive them to the equator of the Bloch sphere. The detunings of the quantum dots relative to the middle dot are chosen to be $[-10, -5, 0, 5, 10]\,\mathrm{ps}^{-1}$. We perform a global optimization search employing a differential evolution algorithm on 35 pulse parameters. We parameterize the phase mask function by splitting it into 32 segments and assigning one parameter to the slope of each segment. In addition to these 32 parameters we also optimize over all three input pulse parameters $\tau$, $\Delta$ and $\Theta$.  To avoid oscillations in time, the phase mask function is smoothed out with a 3rd order spline. We expect that a short $\pi/2$-pulse will successfully drive the states of the quantum dots close to the equator of the Bloch sphere. This is because the shorter the pulse is, the broader is its frequency distribution, leading to a suppressed detuning dependency. For the optimization we use a differential evolution with a population size of eight parameter sets per dimension, where we set one element of the initial population to a simple $100\,\mathrm{fs}$ $\pi/2$-pulse and chose the rest randomly.

The differential evolution algorithm employed 10\,400 ensemble simulations, which each entailed the computation of the full non-Markovain dynamics of 5 independent quantum dots. Using the PT-TEMPO method on all four cores of an Intel i7 processor this took only about 11 hours, while the same computation would have taken more than a month with the conventional TEMPO method. The result of this optimization is shown in the Figs.~\ref{fig:ensemble}b,~\ref{fig:ensemble}e~and~\ref{fig:ensemble}f. Surprisingly, the algorithm found an unexpected pulse form that leads to a root mean square (RMS) distance to the equator of the Bloch sphere of 0.10, which is significantly better than the performance of a $\pi/2$-pulse with the same initial pulse duration of $\tau=245\,\mathrm{fs}$ (see Figs.~\ref{fig:ensemble}c~and~\ref{fig:ensemble}d). Also, it performs slightly better than the shortest $\pi/2$-pulse with $\tau = 30\,\mathrm{fs}$, which yields a RMS distance of 0.12. However, we note that, unlike the $\pi/2$-pulse, the performance of the optimized pulse is sensitively dependent on the exact detuning of the individual quantum dots.


\textit{Conclusion}---
We have shown that the PT-TEMPO method makes optimal control of non-Markovian open quantum systems a feasible task. It is applicable to small systems that couple to a structured bosonic environment and it performs well if the environment correlation function is smooth and decays to zero within some finite time. 
The key idea of the method is to modify the contraction order of the TEMPO tensor network such that the result of the bulk of the computation---corresponding to the contraction of the Feynman-Vernon influence functionals---can be stored and reused for each new trial system Hamiltonian. Finally, we note that this idea can be also applied to other tensor network methods~\cite{Prior2010, Chin2010, Tamascelli2018, Brenes2019, Somoza2019}, opening the way to a family of efficient methods to design quantum control procedures for non-Markovian open quantum systems.

We thank F. Pollock for insightful discussions on the process tensor framework.
G.E.F. acknowledges support from \review{EPSRC (EP/L015110/1).} B.W.L. and J.K. acknowledge support from EPSRC (EP/T014032/1). E.B. acknowledges support from the Irish Research Council (GOIPG/2019/1871)\review{, and} P.R.E. acknowledges support from Science Foundation Ireland (15/IACA/3402).


%



\begin{thebibliography}{51}%
	\makeatletter
	\providecommand \@ifxundefined [1]{%
		\@ifx{#1\undefined}
	}%
	\providecommand \@ifnum [1]{%
		\ifnum #1\expandafter \@firstoftwo
		\else \expandafter \@secondoftwo
		\fi
	}%
	\providecommand \@ifx [1]{%
		\ifx #1\expandafter \@firstoftwo
		\else \expandafter \@secondoftwo
		\fi
	}%
	\providecommand \natexlab [1]{#1}%
	\providecommand \enquote  [1]{``#1''}%
	\providecommand \bibnamefont  [1]{#1}%
	\providecommand \bibfnamefont [1]{#1}%
	\providecommand \citenamefont [1]{#1}%
	\providecommand \href@noop [0]{\@secondoftwo}%
	\providecommand \href [0]{\begingroup \@sanitize@url \@href}%
	\providecommand \@href[1]{\@@startlink{#1}\@@href}%
	\providecommand \@@href[1]{\endgroup#1\@@endlink}%
	\providecommand \@sanitize@url [0]{\catcode `\\12\catcode `\$12\catcode
		`\&12\catcode `\#12\catcode `\^12\catcode `\_12\catcode `\%12\relax}%
	\providecommand \@@startlink[1]{}%
	\providecommand \@@endlink[0]{}%
	\providecommand \url  [0]{\begingroup\@sanitize@url \@url }%
	\providecommand \@url [1]{\endgroup\@href {#1}{\urlprefix }}%
	\providecommand \urlprefix  [0]{URL }%
	\providecommand \Eprint [0]{\href }%
	\providecommand \doibase [0]{https://doi.org/}%
	\providecommand \selectlanguage [0]{\@gobble}%
	\providecommand \bibinfo  [0]{\@secondoftwo}%
	\providecommand \bibfield  [0]{\@secondoftwo}%
	\providecommand \translation [1]{[#1]}%
	\providecommand \BibitemOpen [0]{}%
	\providecommand \bibitemStop [0]{}%
	\providecommand \bibitemNoStop [0]{.\EOS\space}%
	\providecommand \EOS [0]{\spacefactor3000\relax}%
	\providecommand \BibitemShut  [1]{\csname bibitem#1\endcsname}%
	\let\auto@bib@innerbib\@empty
	\bibitem [{\citenamefont {Shapiro}\ and\ \citenamefont
		{Brumer}(2011)}]{Shapiro2012}%
	\BibitemOpen
	\bibfield  {author} {\bibinfo {author} {\bibfnamefont {M.}~\bibnamefont
			{Shapiro}}\ and\ \bibinfo {author} {\bibfnamefont {P.}~\bibnamefont
			{Brumer}},\ }\href {https://doi.org/10.1002/9783527639700} {\emph {\bibinfo
			{title} {Quantum Control of Molecular Processes: Second Edition}}}\ (\bibinfo
	{publisher} {Wiley-VCH Verlag GmbH \& Co. KGaA},\ \bibinfo {address}
	{Weinheim, Germany},\ \bibinfo {year} {2011})\BibitemShut {NoStop}%
	\bibitem [{\citenamefont {Torrontegui}\ \emph {et~al.}(2013)\citenamefont
		{Torrontegui}, \citenamefont {Ibáñez}, \citenamefont {Martínez-Garaot},
		\citenamefont {Modugno}, \citenamefont {{del Campo}}, \citenamefont
		{Guéry-Odelin}, \citenamefont {Ruschhaupt}, \citenamefont {Chen},\ and\
		\citenamefont {Muga}}]{Torrontegui2013}%
	\BibitemOpen
	\bibfield  {author} {\bibinfo {author} {\bibfnamefont {E.}~\bibnamefont
			{Torrontegui}}, \bibinfo {author} {\bibfnamefont {S.}~\bibnamefont
			{Ibáñez}}, \bibinfo {author} {\bibfnamefont {S.}~\bibnamefont
			{Martínez-Garaot}}, \bibinfo {author} {\bibfnamefont {M.}~\bibnamefont
			{Modugno}}, \bibinfo {author} {\bibfnamefont {A.}~\bibnamefont {{del
					Campo}}}, \bibinfo {author} {\bibfnamefont {D.}~\bibnamefont
			{Guéry-Odelin}}, \bibinfo {author} {\bibfnamefont {A.}~\bibnamefont
			{Ruschhaupt}}, \bibinfo {author} {\bibfnamefont {X.}~\bibnamefont {Chen}},\
		and\ \bibinfo {author} {\bibfnamefont {J.~G.}\ \bibnamefont {Muga}},\
	}\bibfield  {title} {\bibinfo {title} {Chapter 2 - shortcuts to
			adiabaticity},\ }in\ \href
	{https://doi.org/https://doi.org/10.1016/B978-0-12-408090-4.00002-5} {\emph
		{\bibinfo {booktitle} {Advances in Atomic, Molecular, and Optical
				Physics}}},\ Vol.~\bibinfo {volume} {62},\ \bibinfo {editor} {edited by\
		\bibinfo {editor} {\bibfnamefont {E.}~\bibnamefont {Arimondo}}, \bibinfo
		{editor} {\bibfnamefont {P.~R.}\ \bibnamefont {Berman}},\ and\ \bibinfo
		{editor} {\bibfnamefont {C.~C.}\ \bibnamefont {Lin}}}\ (\bibinfo  {publisher}
	{Academic Press},\ \bibinfo {year} {2013})\ pp.\ \bibinfo {pages} {117 --
		169}\BibitemShut {NoStop}%
	\bibitem [{\citenamefont {Glaser}\ \emph {et~al.}(2015)\citenamefont {Glaser},
		\citenamefont {Boscain}, \citenamefont {Calarco}, \citenamefont {Koch},
		\citenamefont {K{\"{o}}ckenberger}, \citenamefont {Kosloff}, \citenamefont
		{Kuprov}, \citenamefont {Luy}, \citenamefont {Schirmer}, \citenamefont
		{Schulte-Herbr{\"{u}}ggen}, \citenamefont {Sugny},\ and\ \citenamefont
		{Wilhelm}}]{Glaser2015}%
	\BibitemOpen
	\bibfield  {author} {\bibinfo {author} {\bibfnamefont {S.~J.}\ \bibnamefont
			{Glaser}}, \bibinfo {author} {\bibfnamefont {U.}~\bibnamefont {Boscain}},
		\bibinfo {author} {\bibfnamefont {T.}~\bibnamefont {Calarco}}, \bibinfo
		{author} {\bibfnamefont {C.~P.}\ \bibnamefont {Koch}}, \bibinfo {author}
		{\bibfnamefont {W.}~\bibnamefont {K{\"{o}}ckenberger}}, \bibinfo {author}
		{\bibfnamefont {R.}~\bibnamefont {Kosloff}}, \bibinfo {author} {\bibfnamefont
			{I.}~\bibnamefont {Kuprov}}, \bibinfo {author} {\bibfnamefont
			{B.}~\bibnamefont {Luy}}, \bibinfo {author} {\bibfnamefont {S.}~\bibnamefont
			{Schirmer}}, \bibinfo {author} {\bibfnamefont {T.}~\bibnamefont
			{Schulte-Herbr{\"{u}}ggen}}, \bibinfo {author} {\bibfnamefont
			{D.}~\bibnamefont {Sugny}},\ and\ \bibinfo {author} {\bibfnamefont {F.~K.}\
			\bibnamefont {Wilhelm}},\ }\bibfield  {title} {\bibinfo {title} {{Training
				Schr{\"{o}}dinger's cat: Quantum optimal control: Strategic report on current
				status, visions and goals for research in Europe}},\ }\href
	{https://doi.org/10.1140/epjd/e2015-60464-1} {\bibfield  {journal} {\bibinfo
			{journal} {Eur. Phys. J. D}\ }\textbf {\bibinfo {volume} {69}},\ \bibinfo
		{pages} {279} (\bibinfo {year} {2015})}\BibitemShut {NoStop}%
	\bibitem [{\citenamefont {Koch}(2016)}]{Koch2016}%
	\BibitemOpen
	\bibfield  {author} {\bibinfo {author} {\bibfnamefont {C.~P.}\ \bibnamefont
			{Koch}},\ }\bibfield  {title} {\bibinfo {title} {{Controlling open quantum
				systems: Tools, achievements, and limitations}},\ }\href
	{https://doi.org/10.1088/0953-8984/28/21/213001} {\bibfield  {journal}
		{\bibinfo  {journal} {J. Phys. Condens. Matter}\ }\textbf {\bibinfo {volume}
			{28}},\ \bibinfo {pages} {213001} (\bibinfo {year} {2016})}\BibitemShut
	{NoStop}%
	\bibitem [{\citenamefont {Koch}\ \emph {et~al.}(2019)\citenamefont {Koch},
		\citenamefont {Lemeshko},\ and\ \citenamefont {Sugny}}]{Koch2019}%
	\BibitemOpen
	\bibfield  {author} {\bibinfo {author} {\bibfnamefont {C.~P.}\ \bibnamefont
			{Koch}}, \bibinfo {author} {\bibfnamefont {M.}~\bibnamefont {Lemeshko}},\
		and\ \bibinfo {author} {\bibfnamefont {D.}~\bibnamefont {Sugny}},\ }\bibfield
	{title} {\bibinfo {title} {{Quantum control of molecular rotation}},\ }\href
	{https://doi.org/10.1103/RevModPhys.91.035005} {\bibfield  {journal}
		{\bibinfo  {journal} {Rev. Mod. Phys.}\ }\textbf {\bibinfo {volume} {91}},\
		\bibinfo {pages} {035005} (\bibinfo {year} {2019})}\BibitemShut {NoStop}%
	\bibitem [{\citenamefont {Breuer}\ and\ \citenamefont
		{Petruccione}(2002)}]{Breuer2002}%
	\BibitemOpen
	\bibfield  {author} {\bibinfo {author} {\bibfnamefont {H.-P.}\ \bibnamefont
			{Breuer}}\ and\ \bibinfo {author} {\bibfnamefont {F.}~\bibnamefont
			{Petruccione}},\ }\href
	{https://doi.org/10.1093/acprof:oso/9780199213900.001.0001} {\emph {\bibinfo
			{title} {{The Theory of Open Quantum Systems}}}}\ (\bibinfo  {publisher}
	{Oxford University Press},\ \bibinfo {year} {2002})\BibitemShut {NoStop}%
	\bibitem [{\citenamefont {Chirolli}\ and\ \citenamefont
		{Burkard}(2008)}]{Chirolli2008}%
	\BibitemOpen
	\bibfield  {author} {\bibinfo {author} {\bibfnamefont {L.}~\bibnamefont
			{Chirolli}}\ and\ \bibinfo {author} {\bibfnamefont {G.}~\bibnamefont
			{Burkard}},\ }\bibfield  {title} {\bibinfo {title} {{Decoherence in
				solid-state qubits}},\ }\href {https://doi.org/10.1080/00018730802218067}
	{\bibfield  {journal} {\bibinfo  {journal} {Adv. Phys.}\ }\textbf {\bibinfo
			{volume} {57}},\ \bibinfo {pages} {225} (\bibinfo {year} {2008})}\BibitemShut
	{NoStop}%
	\bibitem [{\citenamefont {Wilson-Rae}\ and\ \citenamefont
		{Imamoğlu}(2002)}]{WilsonRae2002}%
	\BibitemOpen
	\bibfield  {author} {\bibinfo {author} {\bibfnamefont {I.}~\bibnamefont
			{Wilson-Rae}}\ and\ \bibinfo {author} {\bibfnamefont {A.}~\bibnamefont
			{Imamoğlu}},\ }\bibfield  {title} {\bibinfo {title} {{Quantum dot cavity-QED
				in the presence of strong electron-phonon interactions}},\ }\href
	{https://doi.org/10.1103/PhysRevB.65.235311} {\bibfield  {journal} {\bibinfo
			{journal} {Phys. Rev. B}\ }\textbf {\bibinfo {volume} {65}},\ \bibinfo
		{pages} {235311} (\bibinfo {year} {2002})}\BibitemShut {NoStop}%
	\bibitem [{\citenamefont {Galland}\ \emph {et~al.}(2008)\citenamefont
		{Galland}, \citenamefont {H{\"{o}}gele}, \citenamefont {T{\"{u}}reci},\ and\
		\citenamefont {Imamoğlu}}]{Galland2008}%
	\BibitemOpen
	\bibfield  {author} {\bibinfo {author} {\bibfnamefont {C.}~\bibnamefont
			{Galland}}, \bibinfo {author} {\bibfnamefont {A.}~\bibnamefont
			{H{\"{o}}gele}}, \bibinfo {author} {\bibfnamefont {H.~E.}\ \bibnamefont
			{T{\"{u}}reci}},\ and\ \bibinfo {author} {\bibfnamefont {A.}~\bibnamefont
			{Imamoğlu}},\ }\bibfield  {title} {\bibinfo {title} {{Non-Markovian
				Decoherence of Localized Nanotube Excitons by Acoustic Phonons}},\ }\href
	{https://doi.org/10.1103/PhysRevLett.101.067402} {\bibfield  {journal}
		{\bibinfo  {journal} {Phys. Rev. Lett.}\ }\textbf {\bibinfo {volume} {101}},\
		\bibinfo {pages} {067402} (\bibinfo {year} {2008})}\BibitemShut {NoStop}%
	\bibitem [{\citenamefont {Ramsay}\ \emph {et~al.}(2010)\citenamefont {Ramsay},
		\citenamefont {Godden}, \citenamefont {Boyle}, \citenamefont {Gauger},
		\citenamefont {Nazir}, \citenamefont {Lovett}, \citenamefont {Fox},\ and\
		\citenamefont {Skolnick}}]{Ramsay2010}%
	\BibitemOpen
	\bibfield  {author} {\bibinfo {author} {\bibfnamefont {A.~J.}\ \bibnamefont
			{Ramsay}}, \bibinfo {author} {\bibfnamefont {T.~M.}\ \bibnamefont {Godden}},
		\bibinfo {author} {\bibfnamefont {S.~J.}\ \bibnamefont {Boyle}}, \bibinfo
		{author} {\bibfnamefont {E.~M.}\ \bibnamefont {Gauger}}, \bibinfo {author}
		{\bibfnamefont {A.}~\bibnamefont {Nazir}}, \bibinfo {author} {\bibfnamefont
			{B.~W.}\ \bibnamefont {Lovett}}, \bibinfo {author} {\bibfnamefont {A.~M.}\
			\bibnamefont {Fox}},\ and\ \bibinfo {author} {\bibfnamefont {M.~S.}\
			\bibnamefont {Skolnick}},\ }\bibfield  {title} {\bibinfo {title}
		{{Phonon-induced Rabi-frequency renormalization of optically driven single
				InGaAs/GaAs quantum dots}},\ }\href
	{https://doi.org/10.1103/PhysRevLett.105.177402} {\bibfield  {journal}
		{\bibinfo  {journal} {Phys. Rev. Lett.}\ }\textbf {\bibinfo {volume} {105}},\
		\bibinfo {pages} {177402} (\bibinfo {year} {2010})}\BibitemShut {NoStop}%
	\bibitem [{\citenamefont {Roy}\ and\ \citenamefont {Hughes}(2011)}]{Roy2011}%
	\BibitemOpen
	\bibfield  {author} {\bibinfo {author} {\bibfnamefont {C.}~\bibnamefont
			{Roy}}\ and\ \bibinfo {author} {\bibfnamefont {S.}~\bibnamefont {Hughes}},\
	}\bibfield  {title} {\bibinfo {title} {{Phonon-Dressed Mollow Triplet in the
				Regime of Cavity Quantum Electrodynamics: Excitation-Induced Dephasing and
				Nonperturbative Cavity Feeding Effects}},\ }\href
	{https://doi.org/10.1103/PhysRevLett.106.247403} {\bibfield  {journal}
		{\bibinfo  {journal} {Phys. Rev. Lett.}\ }\textbf {\bibinfo {volume} {106}},\
		\bibinfo {pages} {247403} (\bibinfo {year} {2011})}\BibitemShut {NoStop}%
	\bibitem [{\citenamefont {L\"uker}\ \emph {et~al.}(2012)\citenamefont
		{L\"uker}, \citenamefont {Gawarecki}, \citenamefont {Reiter}, \citenamefont
		{Grodecka-Grad}, \citenamefont {Axt}, \citenamefont {Machnikowski},\ and\
		\citenamefont {Kuhn}}]{Luker2012}%
	\BibitemOpen
	\bibfield  {author} {\bibinfo {author} {\bibfnamefont {S.}~\bibnamefont
			{L\"uker}}, \bibinfo {author} {\bibfnamefont {K.}~\bibnamefont {Gawarecki}},
		\bibinfo {author} {\bibfnamefont {D.~E.}\ \bibnamefont {Reiter}}, \bibinfo
		{author} {\bibfnamefont {A.}~\bibnamefont {Grodecka-Grad}}, \bibinfo {author}
		{\bibfnamefont {V.~M.}\ \bibnamefont {Axt}}, \bibinfo {author} {\bibfnamefont
			{P.}~\bibnamefont {Machnikowski}},\ and\ \bibinfo {author} {\bibfnamefont
			{T.}~\bibnamefont {Kuhn}},\ }\bibfield  {title} {\bibinfo {title} {Influence
			of acoustic phonons on the optical control of quantum dots driven by
			adiabatic rapid passage},\ }\href
	{https://doi.org/10.1103/PhysRevB.85.121302} {\bibfield  {journal} {\bibinfo
			{journal} {Phys. Rev. B}\ }\textbf {\bibinfo {volume} {85}},\ \bibinfo
		{pages} {121302} (\bibinfo {year} {2012})}\BibitemShut {NoStop}%
	\bibitem [{\citenamefont {Rebentrost}\ \emph {et~al.}(2009)\citenamefont
		{Rebentrost}, \citenamefont {Serban}, \citenamefont
		{Schulte-Herbr{\"{u}}ggen},\ and\ \citenamefont {Wilhelm}}]{Rebentrost2009}%
	\BibitemOpen
	\bibfield  {author} {\bibinfo {author} {\bibfnamefont {P.}~\bibnamefont
			{Rebentrost}}, \bibinfo {author} {\bibfnamefont {I.}~\bibnamefont {Serban}},
		\bibinfo {author} {\bibfnamefont {T.}~\bibnamefont
			{Schulte-Herbr{\"{u}}ggen}},\ and\ \bibinfo {author} {\bibfnamefont {F.~K.}\
			\bibnamefont {Wilhelm}},\ }\bibfield  {title} {\bibinfo {title} {{Optimal
				control of a qubit coupled to a non-Markovian environment}},\ }\href
	{https://doi.org/10.1103/PhysRevLett.102.090401} {\bibfield  {journal}
		{\bibinfo  {journal} {Phys. Rev. Lett.}\ }\textbf {\bibinfo {volume} {102}},\
		\bibinfo {pages} {090401} (\bibinfo {year} {2009})}\BibitemShut {NoStop}%
	\bibitem [{\citenamefont {Schmidt}\ \emph {et~al.}(2011)\citenamefont
		{Schmidt}, \citenamefont {Negretti}, \citenamefont {Ankerhold}, \citenamefont
		{Calarco},\ and\ \citenamefont {Stockburger}}]{Schmidt2011}%
	\BibitemOpen
	\bibfield  {author} {\bibinfo {author} {\bibfnamefont {R.}~\bibnamefont
			{Schmidt}}, \bibinfo {author} {\bibfnamefont {A.}~\bibnamefont {Negretti}},
		\bibinfo {author} {\bibfnamefont {J.}~\bibnamefont {Ankerhold}}, \bibinfo
		{author} {\bibfnamefont {T.}~\bibnamefont {Calarco}},\ and\ \bibinfo {author}
		{\bibfnamefont {J.~T.}\ \bibnamefont {Stockburger}},\ }\bibfield  {title}
	{\bibinfo {title} {{Optimal control of open quantum systems: Cooperative
				effects of driving and dissipation}},\ }\href
	{https://doi.org/10.1103/PhysRevLett.107.130404} {\bibfield  {journal}
		{\bibinfo  {journal} {Phys. Rev. Lett.}\ }\textbf {\bibinfo {volume} {107}},\
		\bibinfo {pages} {130404} (\bibinfo {year} {2011})}\BibitemShut {NoStop}%
	\bibitem [{\citenamefont {Hwang}\ and\ \citenamefont {Goan}(2012)}]{Hwang2012}%
	\BibitemOpen
	\bibfield  {author} {\bibinfo {author} {\bibfnamefont {B.}~\bibnamefont
			{Hwang}}\ and\ \bibinfo {author} {\bibfnamefont {H.~S.}\ \bibnamefont
			{Goan}},\ }\bibfield  {title} {\bibinfo {title} {{Optimal control for
				non-Markovian open quantum systems}},\ }\href
	{https://doi.org/10.1103/PhysRevA.85.032321} {\bibfield  {journal} {\bibinfo
			{journal} {Phys. Rev. A}\ }\textbf {\bibinfo {volume} {85}},\ \bibinfo
		{pages} {032321} (\bibinfo {year} {2012})}\BibitemShut {NoStop}%
	\bibitem [{\citenamefont {Floether}\ \emph {et~al.}(2012)\citenamefont
		{Floether}, \citenamefont {{De Fouquieres}},\ and\ \citenamefont
		{Schirmer}}]{Floether2012}%
	\BibitemOpen
	\bibfield  {author} {\bibinfo {author} {\bibfnamefont {F.~F.}\ \bibnamefont
			{Floether}}, \bibinfo {author} {\bibfnamefont {P.}~\bibnamefont {{De
					Fouquieres}}},\ and\ \bibinfo {author} {\bibfnamefont {S.~G.}\ \bibnamefont
			{Schirmer}},\ }\bibfield  {title} {\bibinfo {title} {{Robust quantum gates
				for open systems via optimal control: Markovian versus non-Markovian
				dynamics}},\ }\href {https://doi.org/10.1088/1367-2630/14/7/073023}
	{\bibfield  {journal} {\bibinfo  {journal} {New J. Phys.}\ }\textbf {\bibinfo
			{volume} {14}},\ \bibinfo {pages} {073023} (\bibinfo {year}
		{2012})}\BibitemShut {NoStop}%
	\bibitem [{\citenamefont {Reich}\ \emph {et~al.}(2015)\citenamefont {Reich},
		\citenamefont {Katz},\ and\ \citenamefont {Koch}}]{Reich2015}%
	\BibitemOpen
	\bibfield  {author} {\bibinfo {author} {\bibfnamefont {D.~M.}\ \bibnamefont
			{Reich}}, \bibinfo {author} {\bibfnamefont {N.}~\bibnamefont {Katz}},\ and\
		\bibinfo {author} {\bibfnamefont {C.~P.}\ \bibnamefont {Koch}},\ }\bibfield
	{title} {\bibinfo {title} {{Exploiting Non-Markovianity for Quantum
				Control}},\ }\href {https://doi.org/10.1038/srep12430} {\bibfield  {journal}
		{\bibinfo  {journal} {Sci. Rep.}\ }\textbf {\bibinfo {volume} {5}},\ \bibinfo
		{pages} {12430} (\bibinfo {year} {2015})}\BibitemShut {NoStop}%
	\bibitem [{\citenamefont {Addis}\ \emph {et~al.}(2016)\citenamefont {Addis},
		\citenamefont {Laine}, \citenamefont {Gneiting},\ and\ \citenamefont
		{Maniscalco}}]{Addis2016}%
	\BibitemOpen
	\bibfield  {author} {\bibinfo {author} {\bibfnamefont {C.}~\bibnamefont
			{Addis}}, \bibinfo {author} {\bibfnamefont {E.~M.}\ \bibnamefont {Laine}},
		\bibinfo {author} {\bibfnamefont {C.}~\bibnamefont {Gneiting}},\ and\
		\bibinfo {author} {\bibfnamefont {S.}~\bibnamefont {Maniscalco}},\ }\bibfield
	{title} {\bibinfo {title} {{Problem of coherent control in non-Markovian
				open quantum systems}},\ }\href {https://doi.org/10.1103/PhysRevA.94.052117}
	{\bibfield  {journal} {\bibinfo  {journal} {Phys. Rev. A}\ }\textbf {\bibinfo
			{volume} {94}},\ \bibinfo {pages} {052117} (\bibinfo {year}
		{2016})}\BibitemShut {NoStop}%
	\bibitem [{\citenamefont {Puthumpally-Joseph}\ \emph
		{et~al.}(2018)\citenamefont {Puthumpally-Joseph}, \citenamefont {Mangaud},
		\citenamefont {Chevet}, \citenamefont {Desouter-Lecomte}, \citenamefont
		{Sugny},\ and\ \citenamefont {Atabek}}]{Puthumpally-Joseph2018}%
	\BibitemOpen
	\bibfield  {author} {\bibinfo {author} {\bibfnamefont {R.}~\bibnamefont
			{Puthumpally-Joseph}}, \bibinfo {author} {\bibfnamefont {E.}~\bibnamefont
			{Mangaud}}, \bibinfo {author} {\bibfnamefont {V.}~\bibnamefont {Chevet}},
		\bibinfo {author} {\bibfnamefont {M.}~\bibnamefont {Desouter-Lecomte}},
		\bibinfo {author} {\bibfnamefont {D.}~\bibnamefont {Sugny}},\ and\ \bibinfo
		{author} {\bibfnamefont {O.}~\bibnamefont {Atabek}},\ }\bibfield  {title}
	{\bibinfo {title} {{Basic mechanisms in the laser control of non-Markovian
				dynamics}},\ }\href {https://doi.org/10.1103/PhysRevA.97.033411} {\bibfield
		{journal} {\bibinfo  {journal} {Phys. Rev. A}\ }\textbf {\bibinfo {volume}
			{97}},\ \bibinfo {pages} {033411} (\bibinfo {year} {2018})}\BibitemShut
	{NoStop}%
	\bibitem [{\citenamefont {Mangaud}\ \emph {et~al.}(2018)\citenamefont
		{Mangaud}, \citenamefont {Puthumpally-Joseph}, \citenamefont {Sugny},
		\citenamefont {Meier}, \citenamefont {Atabek},\ and\ \citenamefont
		{Desouter-Lecomte}}]{Mangaud2018}%
	\BibitemOpen
	\bibfield  {author} {\bibinfo {author} {\bibfnamefont {E.}~\bibnamefont
			{Mangaud}}, \bibinfo {author} {\bibfnamefont {R.}~\bibnamefont
			{Puthumpally-Joseph}}, \bibinfo {author} {\bibfnamefont {D.}~\bibnamefont
			{Sugny}}, \bibinfo {author} {\bibfnamefont {C.}~\bibnamefont {Meier}},
		\bibinfo {author} {\bibfnamefont {O.}~\bibnamefont {Atabek}},\ and\ \bibinfo
		{author} {\bibfnamefont {M.}~\bibnamefont {Desouter-Lecomte}},\ }\bibfield
	{title} {\bibinfo {title} {{Non-markovianity in the optimal control of an
				open quantum system described by hierarchical equations of motion}},\ }\href
	{https://doi.org/10.1088/1367-2630/aab651} {\bibfield  {journal} {\bibinfo
			{journal} {New J. Phys.}\ }\textbf {\bibinfo {volume} {20}},\ \bibinfo
		{pages} {043050} (\bibinfo {year} {2018})}\BibitemShut {NoStop}%
	\bibitem [{\citenamefont {Goerz}\ and\ \citenamefont
		{Jacobs}(2018)}]{Goerz2018}%
	\BibitemOpen
	\bibfield  {author} {\bibinfo {author} {\bibfnamefont {M.~H.}\ \bibnamefont
			{Goerz}}\ and\ \bibinfo {author} {\bibfnamefont {K.}~\bibnamefont {Jacobs}},\
	}\bibfield  {title} {\bibinfo {title} {{Efficient optimization of state
				preparation in quantum networks using quantum trajectories}},\ }\href
	{https://doi.org/10.1088/2058-9565/aace16} {\bibfield  {journal} {\bibinfo
			{journal} {Quantum Sci. Technol.}\ }\textbf {\bibinfo {volume} {3}},\
		\bibinfo {pages} {045005} (\bibinfo {year} {2018})}\BibitemShut {NoStop}%
	\bibitem [{\citenamefont {Fischer}\ \emph {et~al.}(2019)\citenamefont
		{Fischer}, \citenamefont {Basilewitsch}, \citenamefont {Koch},\ and\
		\citenamefont {Sugny}}]{Fischer2019}%
	\BibitemOpen
	\bibfield  {author} {\bibinfo {author} {\bibfnamefont {J.}~\bibnamefont
			{Fischer}}, \bibinfo {author} {\bibfnamefont {D.}~\bibnamefont
			{Basilewitsch}}, \bibinfo {author} {\bibfnamefont {C.~P.}\ \bibnamefont
			{Koch}},\ and\ \bibinfo {author} {\bibfnamefont {D.}~\bibnamefont {Sugny}},\
	}\bibfield  {title} {\bibinfo {title} {{Time-optimal control of the
				purification of a qubit in contact with a structured environment}},\ }\href
	{https://doi.org/10.1103/PhysRevA.99.033410} {\bibfield  {journal} {\bibinfo
			{journal} {Phys. Rev. A}\ }\textbf {\bibinfo {volume} {99}},\ \bibinfo
		{pages} {033410} (\bibinfo {year} {2019})}\BibitemShut {NoStop}%
	\bibitem [{\citenamefont {Mirkin}\ \emph {et~al.}(2019)\citenamefont {Mirkin},
		\citenamefont {Poggi},\ and\ \citenamefont {Wisniacki}}]{Mirkin2019}%
	\BibitemOpen
	\bibfield  {author} {\bibinfo {author} {\bibfnamefont {N.}~\bibnamefont
			{Mirkin}}, \bibinfo {author} {\bibfnamefont {P.}~\bibnamefont {Poggi}},\ and\
		\bibinfo {author} {\bibfnamefont {D.}~\bibnamefont {Wisniacki}},\ }\bibfield
	{title} {\bibinfo {title} {{Entangling protocols due to non-Markovian
				dynamics}},\ }\href {https://doi.org/10.1103/PhysRevA.99.020301} {\bibfield
		{journal} {\bibinfo  {journal} {Phys. Rev. A}\ }\textbf {\bibinfo {volume}
			{99}},\ \bibinfo {pages} {020301} (\bibinfo {year} {2019})}\BibitemShut
	{NoStop}%
	\bibitem [{\citenamefont {Alipour}\ \emph {et~al.}(2020)\citenamefont
		{Alipour}, \citenamefont {Chenu}, \citenamefont {Rezakhani},\ and\
		\citenamefont {del Campo}}]{Alipour2020}%
	\BibitemOpen
	\bibfield  {author} {\bibinfo {author} {\bibfnamefont {S.}~\bibnamefont
			{Alipour}}, \bibinfo {author} {\bibfnamefont {A.}~\bibnamefont {Chenu}},
		\bibinfo {author} {\bibfnamefont {A.~T.}\ \bibnamefont {Rezakhani}},\ and\
		\bibinfo {author} {\bibfnamefont {A.}~\bibnamefont {del Campo}},\ }\bibfield
	{title} {\bibinfo {title} {Shortcuts to {A}diabaticity in {D}riven {O}pen
			{Q}uantum {S}ystems: {B}alanced {G}ain and {L}oss and {N}on-{M}arkovian
			{E}volution},\ }\href {https://doi.org/10.22331/q-2020-09-28-336} {\bibfield
		{journal} {\bibinfo  {journal} {{Quantum}}\ }\textbf {\bibinfo {volume}
			{4}},\ \bibinfo {pages} {336} (\bibinfo {year} {2020})}\BibitemShut {NoStop}%
	\bibitem [{\citenamefont {Tanimura}\ and\ \citenamefont
		{Kubo}(1989)}]{Tanimura1989}%
	\BibitemOpen
	\bibfield  {author} {\bibinfo {author} {\bibfnamefont {Y.}~\bibnamefont
			{Tanimura}}\ and\ \bibinfo {author} {\bibfnamefont {R.}~\bibnamefont
			{Kubo}},\ }\bibfield  {title} {\bibinfo {title} {{Time Evolution of a Quantum
				System in Contact with a Nearly Gaussian-Markoffian Noise Bath}},\ }\href
	{https://doi.org/10.1143/JPSJ.58.101} {\bibfield  {journal} {\bibinfo
			{journal} {J. Phys. Soc. Japan}\ }\textbf {\bibinfo {volume} {58}},\ \bibinfo
		{pages} {101} (\bibinfo {year} {1989})}\BibitemShut {NoStop}%
	\bibitem [{\citenamefont {Makri}\ and\ \citenamefont
		{Makarov}(1995{\natexlab{a}})}]{Makri1995}%
	\BibitemOpen
	\bibfield  {author} {\bibinfo {author} {\bibfnamefont {N.}~\bibnamefont
			{Makri}}\ and\ \bibinfo {author} {\bibfnamefont {D.~E.}\ \bibnamefont
			{Makarov}},\ }\bibfield  {title} {\bibinfo {title} {{Tensor propagator for
				iterative quantum time evolution of reduced density matrices. I. Theory}},\
	}\href {https://doi.org/10.1063/1.469508} {\bibfield  {journal} {\bibinfo
			{journal} {J. Chem. Phys.}\ }\textbf {\bibinfo {volume} {102}},\ \bibinfo
		{pages} {4600} (\bibinfo {year} {1995}{\natexlab{a}})}\BibitemShut {NoStop}%
	\bibitem [{\citenamefont {Makri}\ and\ \citenamefont
		{Makarov}(1995{\natexlab{b}})}]{Makri1995a}%
	\BibitemOpen
	\bibfield  {author} {\bibinfo {author} {\bibfnamefont {N.}~\bibnamefont
			{Makri}}\ and\ \bibinfo {author} {\bibfnamefont {D.~E.}\ \bibnamefont
			{Makarov}},\ }\bibfield  {title} {\bibinfo {title} {{Tensor propagator for
				iterative quantum time evolution of reduced density matrices. II. Numerical
				methodology}},\ }\href {https://doi.org/10.1063/1.469509} {\bibfield
		{journal} {\bibinfo  {journal} {J. Chem. Phys.}\ }\textbf {\bibinfo {volume}
			{102}},\ \bibinfo {pages} {4611} (\bibinfo {year}
		{1995}{\natexlab{b}})}\BibitemShut {NoStop}%
	\bibitem [{\citenamefont {Prior}\ \emph {et~al.}(2010)\citenamefont {Prior},
		\citenamefont {Chin}, \citenamefont {Huelga},\ and\ \citenamefont
		{Plenio}}]{Prior2010}%
	\BibitemOpen
	\bibfield  {author} {\bibinfo {author} {\bibfnamefont {J.}~\bibnamefont
			{Prior}}, \bibinfo {author} {\bibfnamefont {A.~W.}\ \bibnamefont {Chin}},
		\bibinfo {author} {\bibfnamefont {S.~F.}\ \bibnamefont {Huelga}},\ and\
		\bibinfo {author} {\bibfnamefont {M.~B.}\ \bibnamefont {Plenio}},\ }\bibfield
	{title} {\bibinfo {title} {{Efficient simulation of strong
				system-environment interactions}},\ }\href
	{https://doi.org/10.1103/PhysRevLett.105.050404} {\bibfield  {journal}
		{\bibinfo  {journal} {Phys. Rev. Lett.}\ }\textbf {\bibinfo {volume} {105}},\
		\bibinfo {pages} {050404} (\bibinfo {year} {2010})}\BibitemShut {NoStop}%
	\bibitem [{\citenamefont {Chin}\ \emph {et~al.}(2010)\citenamefont {Chin},
		\citenamefont {Rivas}, \citenamefont {Huelga},\ and\ \citenamefont
		{Plenio}}]{Chin2010}%
	\BibitemOpen
	\bibfield  {author} {\bibinfo {author} {\bibfnamefont {A.~W.}\ \bibnamefont
			{Chin}}, \bibinfo {author} {\bibfnamefont {{\'{A}}.}~\bibnamefont {Rivas}},
		\bibinfo {author} {\bibfnamefont {S.~F.}\ \bibnamefont {Huelga}},\ and\
		\bibinfo {author} {\bibfnamefont {M.~B.}\ \bibnamefont {Plenio}},\ }\bibfield
	{title} {\bibinfo {title} {{Exact mapping between system-reservoir quantum
				models and semi-infinite discrete chains using orthogonal polynomials}},\
	}\href {https://doi.org/10.1063/1.3490188} {\bibfield  {journal} {\bibinfo
			{journal} {J. Math. Phys.}\ }\textbf {\bibinfo {volume} {51}},\ \bibinfo
		{pages} {092109} (\bibinfo {year} {2010})}\BibitemShut {NoStop}%
	\bibitem [{\citenamefont {Cerrillo}\ and\ \citenamefont
		{Cao}(2014)}]{Cerrillo2014}%
	\BibitemOpen
	\bibfield  {author} {\bibinfo {author} {\bibfnamefont {J.}~\bibnamefont
			{Cerrillo}}\ and\ \bibinfo {author} {\bibfnamefont {J.}~\bibnamefont {Cao}},\
	}\bibfield  {title} {\bibinfo {title} {{Non-Markovian dynamical maps:
				Numerical processing of open quantum trajectories}},\ }\href
	{https://doi.org/10.1103/PhysRevLett.112.110401} {\bibfield  {journal}
		{\bibinfo  {journal} {Phys. Rev. Lett.}\ }\textbf {\bibinfo {volume} {112}},\
		\bibinfo {pages} {110401} (\bibinfo {year} {2014})}\BibitemShut {NoStop}%
	\bibitem [{\citenamefont {Iles-Smith}\ \emph {et~al.}(2014)\citenamefont
		{Iles-Smith}, \citenamefont {Lambert},\ and\ \citenamefont
		{Nazir}}]{Iles-Smith2014}%
	\BibitemOpen
	\bibfield  {author} {\bibinfo {author} {\bibfnamefont {J.}~\bibnamefont
			{Iles-Smith}}, \bibinfo {author} {\bibfnamefont {N.}~\bibnamefont
			{Lambert}},\ and\ \bibinfo {author} {\bibfnamefont {A.}~\bibnamefont
			{Nazir}},\ }\bibfield  {title} {\bibinfo {title} {{Environmental dynamics,
				correlations, and the emergence of noncanonical equilibrium states in open
				quantum systems}},\ }\href {https://doi.org/10.1103/PhysRevA.90.032114}
	{\bibfield  {journal} {\bibinfo  {journal} {Phys. Rev. A}\ }\textbf {\bibinfo
			{volume} {90}},\ \bibinfo {pages} {032114} (\bibinfo {year}
		{2014})}\BibitemShut {NoStop}%
	\bibitem [{\citenamefont {Tamascelli}\ \emph {et~al.}(2018)\citenamefont
		{Tamascelli}, \citenamefont {Smirne}, \citenamefont {Huelga},\ and\
		\citenamefont {Plenio}}]{Tamascelli2017}%
	\BibitemOpen
	\bibfield  {author} {\bibinfo {author} {\bibfnamefont {D.}~\bibnamefont
			{Tamascelli}}, \bibinfo {author} {\bibfnamefont {A.}~\bibnamefont {Smirne}},
		\bibinfo {author} {\bibfnamefont {S.~F.}\ \bibnamefont {Huelga}},\ and\
		\bibinfo {author} {\bibfnamefont {M.~B.}\ \bibnamefont {Plenio}},\ }\bibfield
	{title} {\bibinfo {title} {{Nonperturbative Treatment of non-Markovian
				Dynamics of Open Quantum Systems}},\ }\href
	{https://doi.org/10.1103/PhysRevLett.120.030402} {\bibfield  {journal}
		{\bibinfo  {journal} {Phys. Rev. Lett.}\ }\textbf {\bibinfo {volume} {120}},\
		\bibinfo {pages} {30402} (\bibinfo {year} {2018})}\BibitemShut {NoStop}%
	\bibitem [{\citenamefont {Tamascelli}\ \emph {et~al.}(2019)\citenamefont
		{Tamascelli}, \citenamefont {Smirne}, \citenamefont {Lim}, \citenamefont
		{Huelga},\ and\ \citenamefont {Plenio}}]{Tamascelli2018}%
	\BibitemOpen
	\bibfield  {author} {\bibinfo {author} {\bibfnamefont {D.}~\bibnamefont
			{Tamascelli}}, \bibinfo {author} {\bibfnamefont {A.}~\bibnamefont {Smirne}},
		\bibinfo {author} {\bibfnamefont {J.}~\bibnamefont {Lim}}, \bibinfo {author}
		{\bibfnamefont {S.~F.}\ \bibnamefont {Huelga}},\ and\ \bibinfo {author}
		{\bibfnamefont {M.~B.}\ \bibnamefont {Plenio}},\ }\bibfield  {title}
	{\bibinfo {title} {{Efficient Simulation of Finite-Temperature Open Quantum
				Systems}},\ }\href {https://doi.org/10.1103/PhysRevLett.123.090402}
	{\bibfield  {journal} {\bibinfo  {journal} {Phys. Rev. Lett.}\ }\textbf
		{\bibinfo {volume} {123}},\ \bibinfo {pages} {90402} (\bibinfo {year}
		{2019})}\BibitemShut {NoStop}%
	\bibitem [{\citenamefont {Somoza}\ \emph {et~al.}(2019)\citenamefont {Somoza},
		\citenamefont {Marty}, \citenamefont {Lim}, \citenamefont {Huelga},\ and\
		\citenamefont {Plenio}}]{Somoza2019}%
	\BibitemOpen
	\bibfield  {author} {\bibinfo {author} {\bibfnamefont {A.~D.}\ \bibnamefont
			{Somoza}}, \bibinfo {author} {\bibfnamefont {O.}~\bibnamefont {Marty}},
		\bibinfo {author} {\bibfnamefont {J.}~\bibnamefont {Lim}}, \bibinfo {author}
		{\bibfnamefont {S.~F.}\ \bibnamefont {Huelga}},\ and\ \bibinfo {author}
		{\bibfnamefont {M.~B.}\ \bibnamefont {Plenio}},\ }\bibfield  {title}
	{\bibinfo {title} {{Dissipation-Assisted Matrix Product Factorization}},\
	}\href {https://doi.org/10.1103/PhysRevLett.123.100502} {\bibfield  {journal}
		{\bibinfo  {journal} {Phys. Rev. Lett.}\ }\textbf {\bibinfo {volume} {123}},\
		\bibinfo {pages} {100502} (\bibinfo {year} {2019})}\BibitemShut {NoStop}%
	\bibitem [{\citenamefont {Mascherpa}\ \emph {et~al.}(2020)\citenamefont
		{Mascherpa}, \citenamefont {Smirne}, \citenamefont {Somoza}, \citenamefont
		{Fern{\'{a}}ndez-Acebal}, \citenamefont {Donadi}, \citenamefont {Tamascelli},
		\citenamefont {Huelga},\ and\ \citenamefont {Plenio}}]{Mascherpa2019}%
	\BibitemOpen
	\bibfield  {author} {\bibinfo {author} {\bibfnamefont {F.}~\bibnamefont
			{Mascherpa}}, \bibinfo {author} {\bibfnamefont {A.}~\bibnamefont {Smirne}},
		\bibinfo {author} {\bibfnamefont {A.~D.}\ \bibnamefont {Somoza}}, \bibinfo
		{author} {\bibfnamefont {P.}~\bibnamefont {Fern{\'{a}}ndez-Acebal}}, \bibinfo
		{author} {\bibfnamefont {S.}~\bibnamefont {Donadi}}, \bibinfo {author}
		{\bibfnamefont {D.}~\bibnamefont {Tamascelli}}, \bibinfo {author}
		{\bibfnamefont {S.~F.}\ \bibnamefont {Huelga}},\ and\ \bibinfo {author}
		{\bibfnamefont {M.~B.}\ \bibnamefont {Plenio}},\ }\bibfield  {title}
	{\bibinfo {title} {{Optimized auxiliary oscillators for the simulation of
				general open quantum systems}},\ }\href
	{https://doi.org/10.1103/PhysRevA.101.052108} {\bibfield  {journal} {\bibinfo
			{journal} {Phys. Rev. A}\ }\textbf {\bibinfo {volume} {101}},\ \bibinfo
		{pages} {052108} (\bibinfo {year} {2020})}\BibitemShut {NoStop}%
	\bibitem [{\citenamefont {Brenes}\ \emph {et~al.}(2020)\citenamefont {Brenes},
		\citenamefont {Mendoza-Arenas}, \citenamefont {Purkayastha}, \citenamefont
		{Mitchison}, \citenamefont {Clark},\ and\ \citenamefont
		{Goold}}]{Brenes2019}%
	\BibitemOpen
	\bibfield  {author} {\bibinfo {author} {\bibfnamefont {M.}~\bibnamefont
			{Brenes}}, \bibinfo {author} {\bibfnamefont {J.~J.}\ \bibnamefont
			{Mendoza-Arenas}}, \bibinfo {author} {\bibfnamefont {A.}~\bibnamefont
			{Purkayastha}}, \bibinfo {author} {\bibfnamefont {M.~T.}\ \bibnamefont
			{Mitchison}}, \bibinfo {author} {\bibfnamefont {S.~R.}\ \bibnamefont
			{Clark}},\ and\ \bibinfo {author} {\bibfnamefont {J.}~\bibnamefont {Goold}},\
	}\bibfield  {title} {\bibinfo {title} {{Tensor-Network Method to Simulate
				Strongly Interacting Quantum Thermal Machines}},\ }\href
	{https://doi.org/10.1103/PhysRevX.10.031040} {\bibfield  {journal} {\bibinfo
			{journal} {Phys. Rev. X}\ }\textbf {\bibinfo {volume} {10}},\ \bibinfo
		{pages} {031040} (\bibinfo {year} {2020})}\BibitemShut {NoStop}%
	\bibitem [{\citenamefont {Tanimura}(2020)}]{Tanimura2020}%
	\BibitemOpen
	\bibfield  {author} {\bibinfo {author} {\bibfnamefont {Y.}~\bibnamefont
			{Tanimura}},\ }\bibfield  {title} {\bibinfo {title} {{Numerically ``exact''
				approach to open quantum dynamics: The hierarchical equations of motion
				(HEOM)}},\ }\href {https://doi.org/10.1063/5.0011599} {\bibfield  {journal}
		{\bibinfo  {journal} {J. Chem. Phys.}\ }\textbf {\bibinfo {volume} {153}},\
		\bibinfo {pages} {020901} (\bibinfo {year} {2020})}\BibitemShut {NoStop}%
	\bibitem [{\citenamefont {{de Vega}}\ and\ \citenamefont
		{Alonso}(2017)}]{DeVega2017}%
	\BibitemOpen
	\bibfield  {author} {\bibinfo {author} {\bibfnamefont {I.}~\bibnamefont {{de
					Vega}}}\ and\ \bibinfo {author} {\bibfnamefont {D.}~\bibnamefont {Alonso}},\
	}\bibfield  {title} {\bibinfo {title} {{Dynamics of non-Markovian open
				quantum systems}},\ }\href {https://doi.org/10.1103/RevModPhys.89.015001}
	{\bibfield  {journal} {\bibinfo  {journal} {Rev. Mod. Phys.}\ }\textbf
		{\bibinfo {volume} {89}},\ \bibinfo {pages} {015001} (\bibinfo {year}
		{2017})}\BibitemShut {NoStop}%
	\bibitem [{\citenamefont {Strathearn}\ \emph {et~al.}(2018)\citenamefont
		{Strathearn}, \citenamefont {Kirton}, \citenamefont {Kilda}, \citenamefont
		{Keeling},\ and\ \citenamefont {Lovett}}]{Strathearn2018}%
	\BibitemOpen
	\bibfield  {author} {\bibinfo {author} {\bibfnamefont {A.}~\bibnamefont
			{Strathearn}}, \bibinfo {author} {\bibfnamefont {P.}~\bibnamefont {Kirton}},
		\bibinfo {author} {\bibfnamefont {D.}~\bibnamefont {Kilda}}, \bibinfo
		{author} {\bibfnamefont {J.}~\bibnamefont {Keeling}},\ and\ \bibinfo {author}
		{\bibfnamefont {B.~W.}\ \bibnamefont {Lovett}},\ }\bibfield  {title}
	{\bibinfo {title} {{Efficient non-Markovian quantum dynamics using
				time-evolving matrix product operators}},\ }\href
	{https://doi.org/10.1038/s41467-018-05617-3} {\bibfield  {journal} {\bibinfo
			{journal} {Nat. Commun.}\ }\textbf {\bibinfo {volume} {9}},\ \bibinfo {pages}
		{3322} (\bibinfo {year} {2018})}\BibitemShut {NoStop}%
	\bibitem [{\citenamefont {Strathearn}(2020)}]{Strathearn2019}%
	\BibitemOpen
	\bibfield  {author} {\bibinfo {author} {\bibfnamefont {A.}~\bibnamefont
			{Strathearn}},\ }\href {https://doi.org/10.1007/978-3-030-54975-6} {\emph
		{\bibinfo {title} {{Modelling Non-Markovian Quantum Systems Using Tensor
					Networks}}}},\ Springer Theses\ (\bibinfo  {publisher} {Springer
		International Publishing},\ \bibinfo {address} {Cham},\ \bibinfo {year}
	{2020})\BibitemShut {NoStop}%
	\bibitem [{\citenamefont {Pollock}\ \emph
		{et~al.}(2018{\natexlab{a}})\citenamefont {Pollock}, \citenamefont
		{Rodr{\'{i}}guez-Rosario}, \citenamefont {Frauenheim}, \citenamefont
		{Paternostro},\ and\ \citenamefont {Modi}}]{Pollock2018}%
	\BibitemOpen
	\bibfield  {author} {\bibinfo {author} {\bibfnamefont {F.~A.}\ \bibnamefont
			{Pollock}}, \bibinfo {author} {\bibfnamefont {C.}~\bibnamefont
			{Rodr{\'{i}}guez-Rosario}}, \bibinfo {author} {\bibfnamefont
			{T.}~\bibnamefont {Frauenheim}}, \bibinfo {author} {\bibfnamefont
			{M.}~\bibnamefont {Paternostro}},\ and\ \bibinfo {author} {\bibfnamefont
			{K.}~\bibnamefont {Modi}},\ }\bibfield  {title} {\bibinfo {title}
		{{Non-Markovian quantum processes: Complete framework and efficient
				characterization}},\ }\href {https://doi.org/10.1103/PhysRevA.97.012127}
	{\bibfield  {journal} {\bibinfo  {journal} {Phys. Rev. A}\ }\textbf {\bibinfo
			{volume} {97}},\ \bibinfo {pages} {012127} (\bibinfo {year}
		{2018}{\natexlab{a}})}\BibitemShut {NoStop}%
	\bibitem [{\citenamefont {{The TEMPO collaboration}}(2020)}]{TimeEvolvingMPO}%
	\BibitemOpen
	\bibfield  {author} {\bibinfo {author} {\bibnamefont {{The TEMPO
					collaboration}}},\ }\href {https://doi.org/10.5281/zenodo.4428316} {\bibinfo
		{title} {{TimeEvolvingMPO: A Python 3 package to efficiently compute
				non-Markovian open quantum systems.}}} (\bibinfo {year} {2020})\BibitemShut
	{NoStop}%
	\bibitem [{\citenamefont {Feynman}\ and\ \citenamefont
		{Vernon}(1963)}]{Feynman1963}%
	\BibitemOpen
	\bibfield  {author} {\bibinfo {author} {\bibfnamefont {R.~P.}\ \bibnamefont
			{Feynman}}\ and\ \bibinfo {author} {\bibfnamefont {F.~L.}\ \bibnamefont
			{Vernon}},\ }\bibfield  {title} {\bibinfo {title} {{The theory of a general
				quantum system interacting with a linear dissipative system}},\ }\href
	{https://doi.org/10.1016/0003-4916(63)90068-X} {\bibfield  {journal}
		{\bibinfo  {journal} {Ann. of Phys.}\ }\textbf {\bibinfo {volume} {24}},\
		\bibinfo {pages} {118} (\bibinfo {year} {1963})}\BibitemShut {NoStop}%
	\bibitem [{\citenamefont {Or{\'{u}}s}(2014)}]{Orus2014}%
	\BibitemOpen
	\bibfield  {author} {\bibinfo {author} {\bibfnamefont {R.}~\bibnamefont
			{Or{\'{u}}s}},\ }\bibfield  {title} {\bibinfo {title} {{A practical
				introduction to tensor networks: Matrix product states and projected
				entangled pair states}},\ }\href {https://doi.org/10.1016/j.aop.2014.06.013}
	{\bibfield  {journal} {\bibinfo  {journal} {Ann. of Phys.}\ }\textbf
		{\bibinfo {volume} {349}},\ \bibinfo {pages} {117} (\bibinfo {year}
		{2014})}\BibitemShut {NoStop}%
	\bibitem [{\citenamefont {Cirac}\ \emph {et~al.}(2020)\citenamefont {Cirac},
		\citenamefont {Perez-Garcia}, \citenamefont {Schuch},\ and\ \citenamefont
		{Verstraete}}]{Cirac2020}%
	\BibitemOpen
	\bibfield  {author} {\bibinfo {author} {\bibfnamefont {I.}~\bibnamefont
			{Cirac}}, \bibinfo {author} {\bibfnamefont {D.}~\bibnamefont {Perez-Garcia}},
		\bibinfo {author} {\bibfnamefont {N.}~\bibnamefont {Schuch}},\ and\ \bibinfo
		{author} {\bibfnamefont {F.}~\bibnamefont {Verstraete}},\ }\bibfield  {title}
	{\bibinfo {title} {Matrix product states and projected entangled pair states:
			Concepts, symmetries, and theorems},\ }\Eprint
	{https://arxiv.org/abs/2011.12127} {2011.12127}  (\bibinfo {year} {2020}),\
	\bibinfo {note} {preprint}\BibitemShut {NoStop}%
	\bibitem [{\citenamefont {J\o{}rgensen}\ and\ \citenamefont
		{Pollock}(2019)}]{Jorgensen2019}%
	\BibitemOpen
	\bibfield  {author} {\bibinfo {author} {\bibfnamefont {M.~R.}\ \bibnamefont
			{J\o{}rgensen}}\ and\ \bibinfo {author} {\bibfnamefont {F.~A.}\ \bibnamefont
			{Pollock}},\ }\bibfield  {title} {\bibinfo {title} {Exploiting the causal
			tensor network structure of quantum processes to efficiently simulate
			non-markovian path integrals},\ }\href
	{https://doi.org/10.1103/PhysRevLett.123.240602} {\bibfield  {journal}
		{\bibinfo  {journal} {Phys. Rev. Lett.}\ }\textbf {\bibinfo {volume} {123}},\
		\bibinfo {pages} {240602} (\bibinfo {year} {2019})}\BibitemShut {NoStop}%
	\bibitem [{\citenamefont {Milz}\ \emph {et~al.}(2019)\citenamefont {Milz},
		\citenamefont {Kim}, \citenamefont {Pollock},\ and\ \citenamefont
		{Modi}}]{Milz2019}%
	\BibitemOpen
	\bibfield  {author} {\bibinfo {author} {\bibfnamefont {S.}~\bibnamefont
			{Milz}}, \bibinfo {author} {\bibfnamefont {M.~S.}\ \bibnamefont {Kim}},
		\bibinfo {author} {\bibfnamefont {F.~A.}\ \bibnamefont {Pollock}},\ and\
		\bibinfo {author} {\bibfnamefont {K.}~\bibnamefont {Modi}},\ }\bibfield
	{title} {\bibinfo {title} {{Completely Positive Divisibility Does Not Mean
				Markovianity}},\ }\href {https://doi.org/10.1103/PhysRevLett.123.040401}
	{\bibfield  {journal} {\bibinfo  {journal} {Phys. Rev. Lett.}\ }\textbf
		{\bibinfo {volume} {123}},\ \bibinfo {pages} {040401} (\bibinfo {year}
		{2019})}\BibitemShut {NoStop}%
	\bibitem [{\citenamefont {Pollock}\ \emph
		{et~al.}(2018{\natexlab{b}})\citenamefont {Pollock}, \citenamefont
		{Rodr{\'{i}}guez-Rosario}, \citenamefont {Frauenheim}, \citenamefont
		{Paternostro},\ and\ \citenamefont {Modi}}]{Pollock2018a}%
	\BibitemOpen
	\bibfield  {author} {\bibinfo {author} {\bibfnamefont {F.~A.}\ \bibnamefont
			{Pollock}}, \bibinfo {author} {\bibfnamefont {C.}~\bibnamefont
			{Rodr{\'{i}}guez-Rosario}}, \bibinfo {author} {\bibfnamefont
			{T.}~\bibnamefont {Frauenheim}}, \bibinfo {author} {\bibfnamefont
			{M.}~\bibnamefont {Paternostro}},\ and\ \bibinfo {author} {\bibfnamefont
			{K.}~\bibnamefont {Modi}},\ }\bibfield  {title} {\bibinfo {title}
		{{Operational Markov Condition for Quantum Processes}},\ }\href
	{https://doi.org/10.1103/PhysRevLett.120.040405} {\bibfield  {journal}
		{\bibinfo  {journal} {Phys. Rev. Lett.}\ }\textbf {\bibinfo {volume} {120}},\
		\bibinfo {pages} {040405} (\bibinfo {year} {2018}{\natexlab{b}})}\BibitemShut
	{NoStop}%
	\bibitem [{\citenamefont {Eastham}\ \emph {et~al.}(2013)\citenamefont
		{Eastham}, \citenamefont {Spracklen},\ and\ \citenamefont
		{Keeling}}]{Eastham2013}%
	\BibitemOpen
	\bibfield  {author} {\bibinfo {author} {\bibfnamefont {P.~R.}\ \bibnamefont
			{Eastham}}, \bibinfo {author} {\bibfnamefont {A.~O.}\ \bibnamefont
			{Spracklen}},\ and\ \bibinfo {author} {\bibfnamefont {J.}~\bibnamefont
			{Keeling}},\ }\bibfield  {title} {\bibinfo {title} {{Lindblad theory of
				dynamical decoherence of quantum-dot excitons}},\ }\href
	{https://doi.org/10.1103/PhysRevB.87.195306} {\bibfield  {journal} {\bibinfo
			{journal} {Phys. Rev. B}\ }\textbf {\bibinfo {volume} {87}},\ \bibinfo
		{pages} {195306} (\bibinfo {year} {2013})}\BibitemShut {NoStop}%
	\bibitem [{\citenamefont {Weiner}\ \emph {et~al.}(1992)\citenamefont {Weiner},
		\citenamefont {Leaird}, \citenamefont {Patel},\ and\ \citenamefont
		{Wullert}}]{Weiner1992}%
	\BibitemOpen
	\bibfield  {author} {\bibinfo {author} {\bibfnamefont {A.~M.}\ \bibnamefont
			{Weiner}}, \bibinfo {author} {\bibfnamefont {D.~E.}\ \bibnamefont {Leaird}},
		\bibinfo {author} {\bibfnamefont {J.~S.}\ \bibnamefont {Patel}},\ and\
		\bibinfo {author} {\bibfnamefont {J.~R.}\ \bibnamefont {Wullert}},\
	}\bibfield  {title} {\bibinfo {title} {{Programmable Shaping of Femtosecond
				Optical Pulses by Use of 128-Element Liquid Crystal Phase Modulator}},\
	}\href {https://doi.org/10.1109/3.135209} {\bibfield  {journal} {\bibinfo
			{journal} {IEEE J. Quantum Electron.}\ }\textbf {\bibinfo {volume} {28}},\
		\bibinfo {pages} {908} (\bibinfo {year} {1992})}\BibitemShut {NoStop}%
	\bibitem [{\citenamefont {Weiner}(2000)}]{Weiner2000}%
	\BibitemOpen
	\bibfield  {author} {\bibinfo {author} {\bibfnamefont {A.~M.}\ \bibnamefont
			{Weiner}},\ }\bibfield  {title} {\bibinfo {title} {{Femtosecond pulse shaping
				using spatial light modulators}},\ }\href {https://doi.org/10.1063/1.1150614}
	{\bibfield  {journal} {\bibinfo  {journal} {Rev. Sci. Instrum.}\ }\textbf
		{\bibinfo {volume} {71}},\ \bibinfo {pages} {1929} (\bibinfo {year}
		{2000})}\BibitemShut {NoStop}%
\end{thebibliography}
\end{document}